**AI and Education: An Investigation into the Use of ChatGPT for Systems Thinking.**

Holger Arndt

Friedrich-Alexander Universität Erlangen-Nürnberg, Germany

**Abstract**

This exploratory study investigates the potential of the artificial intelligence tool, ChatGPT, to support systems thinking (ST) in various subjects. Using both general and subject-specific prompts, the study assesses the accuracy, helpfulness, and reliability of ChatGPT's responses across different versions of the tool. The results indicate that ChatGPT can provide largely correct and very helpful responses in various subjects, demonstrating its potential as a tool for enhancing ST skills. However, occasional inaccuracies highlight the need for users to remain critical of ChatGPT's responses. Despite some limitations, this study suggests that with careful use and attention to its idiosyncrasies, ChatGPT can be a valuable tool for teaching and learning ST.

**Introduction**

In today's increasingly complex world, systems thinking (ST) emerges as an invaluable skill to equip our students with. It fosters a broader perspective, encouraging individuals to recognize the interconnectedness and complexity of various phenomena, thereby enhancing their understanding of the world and enabling more effective actions (Binkley et al., 2012; Yoon et al., 2017).

Complex situations, ranging from ecological problems to economic issues, social relationships, and health concerns, often confront even children. Successfully navigating these situations requires managing various aspects: perception, evaluation, understanding, consideration of alternatives, decision-making, taking action, and reflection. Children often develop their own explanations and build knowledge from real-life experiences, even in the absence of formal education (Arndt & Kopp, 2017). However, these everyday understandings risk evolving into oversimplified mental models or flawed misconceptions (Berti & Bombi, 1988; diSessa, 2006; Duit, 1999; Norman, 1983). To mitigate this, it becomes crucial to integrate systems thinking into formal education, with educators playing a key role in fostering these skills among students. Numerous studies have underscored the importance and benefits of promoting systems thinking abilities through formal education, including in K-12 curricula and higher education (Frank & Kordova, 2013; Pavlov et al., 2014; Richmond, 1991; Wheat, 2008).

Before exploring ways to enhance ST, it is critical to refine the concept given the diversity of definitions present in the literature (Forrester, 1994; Richmond, 1994; Senge, 1990; Sweeney and Sterman, 2000; Stave and Hopper, 2007; Squires et al., 2011). Arnold and Wade (2015) synthesized these definitions into a comprehensive one: "Systems thinking is a set of synergistic analytic skills used to improve the capability of identifying and understanding systems, predicting their behaviors, and devising modifications to them in order to produce desired effects. These skills work together as a system." They further elaborate that recognizing interconnections, understanding feedback and system structure, differentiating types of stocks, flows, and variables, identifying and understanding non-linear relationships, understanding dynamic behavior, reducing complexity by modeling systems conceptually, and understanding systems at different scales are all fundamental elements of ST.

When considering how to improve ST skills, theoretical literature and empirical studies typically point to System Dynamics (SD) and causal loop diagrams (CLD) (Arndt, 2006; Hovmand & O'Sullivan, 2008; Hung, 2008; Larsson, 2009; Sterman, 2000). While students can relatively easily understand CLDs, SD modeling demands significantly more time and expertise, a factor which may impede its broader implementation in classrooms. However, the advent of artificial intelligence, especially in the form of language models like ChatGPT, offers promising potential in the education sector. This article primarily investigates the potential of ChatGPT in fostering systems thinking among students.

The remainder of the article is structured as follows: Section 2 offers background information on ChatGPT and its significance in learning. Section 3 presents the research questions guiding this study. Section 4 outlines the methodology. Finally, the results are presented and discussed in the concluding sections.

**The Role of ChatGPT in Learning**

In recent years, advancements in AI technologies have permeated numerous sectors, including education (Kasneci et al., 2023; Mogavi et al., 2023). Natural Language Processing is a significant development within AI, this equips computers with the capacity to interpret and generate human language, as a result facilitating human-computer communication. A key technique within Natural Language Processing, large language models, are trained on extensive data sets, enabling them to learn patterns, context, and grammar to produce coherent responses.

A prominent example is ChatGPT, launched by OpenAI in November 2022. This tool utilizes the Generative Pre-trained Transformer (GPT) model, with the freely available version operating on GPT3.5. As of March 2023, an enhanced version, GPT-4, has been available to paid users. This updated version, trained on a larger text corpus, generally provides higher quality responses and supports the integration of plugins to extend ChatGPT's capabilities, such as creating visualizations and performing calculations.

ChatGPT is able to facilitate personalized learning experiences, a critical asset for acquiring systems thinking skills. It offers immediate feedback, promotes self-paced learning, and engages students in interactive learning experiences (Farrokhnia et al., 2023). Given the heterogeneity within classrooms, ChatGPT's capacity to customize responses to individual queries presents a significant advantage. By providing explanations, addressing questions, engaging in dialogues, providing examples, and explaining a topic's relevance, ChatGPT assists learners in constructing knowledge and comprehension. This interactive style encourages students to ask questions they might otherwise hesitate to express in front of their peers, receive constructive feedback, and review previous conversations, fostering a continuous learning experience.

Despite these promising opportunities, using ChatGPT in teaching and learning contexts also introduces potential challenges that require careful consideration. Technical limitations, although likely to be addressed in the future, currently restrict ChatGPT's inputs and outputs to text, preventing it from processing images or logical diagrams. However, plugins such as "Show Me" and "Code Interpreter" already offer some visualization capabilities. The basic version of ChatGPT only contains information up until September 2021. However, plugins such as "WebPilot" enable it to access current information from the web.

Data security is another concern, as user prompts could potentially be used for further model training. Moreover, there is a risk of ChatGPT being used for cheating, such as completing students' homework, a concern that highlights the need for reevaluating homework and assessment tasks (Adiguzel et al., 2023).

Over-reliance on AI for learning also presents a risk. While AI can enhance learning experiences and offer individualized support, it should not replace the crucial role of teachers or undermine the value of human interaction in learning environments. Teachers contribute far more than knowledge transfer; they provide context, foster critical thinking, and promote socio-emotional learning (Farrokhnia et al., 2023).

The inherent variability in ChatGPT's responses to identical prompts means different students may receive different responses, potentially leading to unequal learning opportunities and fairness issues. This is especially the case when learning with ChatGPT occurs at different times, as ChatGPT's answer quality can vary significantly within short timeframes of only a couple of months (Chen et al., 2023).

Perhaps most critically, it is vital to remember that ChatGPT doesn't truly "understand" the users' prompts or its own responses. It derives its "understanding" from a vast amount of text without reflecting on the quality of these sources. This lack of discernment could lead to the reproduction of problematic information, biased or discriminatory responses, and incorrect answers (Cooper, 2023). For students and their teachers, it may be challenging to assess the accuracy of the answers, potentially leading to the assimilation of incorrect information or the development of misconceptions. This risk is particularly high for complex and abstract topics, such as those within systems thinking, where errors may not only be more likely but also harder to detect.

**Research Questions**

Given the potential of ChatGPT as a personalized tutor, providing individualized learning experiences and potentially fostering systems thinking (ST) skills, it must be evaluated whether its benefits outweigh the challenges. This consideration prompts the following research questions:

RQ 1: To what extent are the answers provided by ChatGPT misleading when used for learning about topics and ST?
- Significance: It is critical to assess the dependability and precision of ChatGPT's responses. It will determine the degree of trust we can place in ChatGPT's answers, the level of teacher supervision required, and the feasibility of self-regulated learning without the risk of fostering misconceptions. This applies to both subject content and ST aspects.

RQ2: How and to what extent can ChatGPT improve the eight ST elements identified by Arnold and Wade (2015): recognizing interconnections, identifying and understanding feedback, understanding system structure, differentiating types of stocks, flows, variables, identifying and understanding non-linear relationships, understanding dynamic behavior, reducing complexity by modeling systems conceptually, and understanding systems at different scales?
- Significance: These elements were synthesized based on a comprehensive set of ST definitions and represent the skills deemed relevant by ST researchers and practitioners. It is crucial to understand which of these can be enhanced by ChatGPT.

RQ3: How and to what extent can ChatGPT facilitate learning about causal loop diagrams (CLDs) and assist in their creation?
- Significance: CLDs are widely considered an important tool for ST and understanding complex systems. Given the importance of CLDs for ST, it is worthwhile to examine ChatGPT's potential in relation to CLDs in detail.

RQ4: How and to what extent can ChatGPT facilitate learning about System Dynamics (SD) and assist in creating SD models?
- Significance: Similar to RQ3 and CLDs, SD is partially encompassed in RQ2 but warrants special focus because of its importance.

RQ5: Are there any significant differences in quality between GPT 3.5, GPT 4, and GPT 4 with plugins when used for acquiring ST skills?
- Significance: It is essential to evaluate any disparities in performance between GPT 3.5 and GPT 4 to assess the potential enhancements offered by the newer version. By identifying significant differences, educators can decide whether the paid version is worth the investment. Also, because the use of plugins can change the answers, it is reasonable to check if there is a significant difference between using GPT4 with and without plugins.

RQ6: Are the answers provided by ChatGPT reliable?
- Significance: Consistency in ChatGPT's responses is crucial to establish trust in its educational value. Variation in the answers between different interactions raises questions regarding the reliability of the information provided. Furthermore, if students in a class receive answers of different quality, this can raise issues of fairness.

By addressing these research questions, this study aims to offer valuable insights into ChatGPT's potential to facilitate learning about complex topics and foster ST skills.

**Methodology**

Given the lack of scientific literature on systems thinking (ST) with natural language processing-AI tools like ChatGPT at the time of this study, an explorative and qualitative approach was adopted to address the research questions.

A) The first step involved testing a broad set of prompts in various ST situations through a trial-and-error process, resulting in a list of potentially promising prompts (see Appendices A and B). These prompts formed the basis for subsequent investigations.

B) To examine general learning about ST, causal loop diagrams (CLDs), and System Dynamics (SD) modelling, ChatGPT was asked to explain relevant concepts and provide examples (RQ1, 2, 3, 4) primarily using the prompts in Appendix A. To evaluate potential differences in ChatGPT's responses based on the GPT version, all prompts were run on both GPT 3.5 and GPT 4. When using GPT 4, it was tested both with and without plugins (RQ5). Each set of prompts within each GPT version was executed twice to assess variations in response quality (RQ6). ChatGPT's responses were evaluated based on three criteria—correctness, helpfulness, and reliability—using a four-point Likert scale.

Two trained raters, familiar with the topics as well as ST, CLDs, and SD modelling, evaluated ChatGPT's responses. To ensure reliability and consistency, the raters underwent a training program to familiarize them with the rating scale and the criteria associated with each category. In the event of divergent ratings, an independent third rater intervened, with the final category determined by majority vote. The inter-rater reliability was assessed using Cohen's Kappa coefficient.

C) Three complex, subject-specific topics were chosen as case studies: spread of diseases (biology focus), macroeconomics (economics), and heat transfer (physics). These topics were selected from different subjects to assess if ChatGPT's performance varies by subject (part of RQ1). ChatGPT was asked to explain these topics with a specific focus on ST elements (RQ2), and assist in creating CLDs (RQ3) and SD models (RQ4).

Similar to Step B, each topic was explored with different versions of ChatGPT to provide more data for RQ5 and to ensure a wide variety of subjects. In addition to GPT 3, GPT 4, and GPT 4 with the "Show Me" plugin, the "code interpreter" tool, released in July 2023, was also used.

The process was highly interactive, mimicking a real learning process. While standardized questions (Appendix B) were used, a significant part of the conversation consisted of individual follow-up questions. Due to the interactive nature of this process and the divergent conversations that ensued, using scales, as in Step B, was deemed inappropriate.

Instead, the conversations were verbally described with a focus on quality and notable findings. The corresponding comments are included in the data collection set.

The data collection primarily took place between May and June 2023, with additional data collected in July 2023 following the release of the "code interpreter" tool. It should be noted that ChatGPT's response quality changes over time because of technological advancements, model retraining, or intentional quality reductions to decrease computing load (Chen et al., 2023). The data collection can be accessed online (https://doi.org/10.5281/zenodo.8177162).

**Results**

The initial phase of the project yielded a list of potentially effective prompts for various systems thinking (ST) situations, compiled through a trial-and-error process. These prompts, which typically generate useful results, are listed in Appendices A and B. In n relation to causal loop diagrams (CLDs) highly detailed prompts were more successful, particularly those that include the polarity of a connection. Furthermore, persistence in repeating prompts occasionally improved outcomes.

Table 1 summarizes the results of the questions concerning general aspects of ST, CLDs, and SD, with Cohen's Kappa indicating an almost perfect inter-rater reliability at 0.811.

| Scale Point | Correct | Helpful | Similar |
|---|---|---|---|
| 1 | 65 | 63 | 26 |
| 2 | 0 | 1 | 38 |
| 3 | 0 | 1 | 2 |
| 4 | 1 | 1 | 0 |

Table 1: Distribution of a 4-point Likert scale in respect to the correctness, helpfulness, and similarity of ChatGPT's answers.

The case study results are documented and commented on in detail in the data file. When answering the research questions, specific page numbers are provided for reference to the corresponding parts.

RQ 1: To what degree are the answers provided by ChatGPT misleading when used for learning about topics and ST?

The standardized questions in Appendix A, focusing on explaining ST concepts and providing examples, were almost always correctly answered by ChatGPT. Notable exceptions included the instances where GPT4 with the "Show Me" plugin was asked to provide an example of a CLD and failed to do so correctly (p. 94f).

Across the three case studies, the accuracy levels showed more variation, with "GPT 4 + Show Me" consistently performing the worst. Inquiries about the topic and its correlation with the 8 identified elements of ST yielded "very good" ratings for all answers (GPT3.5, GPT4, and GPT) or a "good rating" for "GPT 4 + Show Me." However, when asked to describe Causal Loop Diagrams (CLDs), occasional mistakes regarding the polarity of the connections emerged. Nevertheless, the verbal explanations provided were almost always correct. Here's a typical example:

> Connect **Outside Temperature** to **Heat Loss to Outside** with a "+" arrow. The colder it is outside, the faster heat will be lost from the home. Therefore, an increase in Outside Temperature leads to a decrease in Heat Loss to Outside, indicating an inverse relationship. (p. 281)

No mistakes with the potential to create misconceptions were identified when supporting the modeling process. Occasional errors were found, mainly concerning detailed instructions on how to use a specific modeling software.

RQ2: How and to what degree can ChatGPT help improving the eight elements of ST according to Arnold and Wade (2015)?

All explanations and examples related to the eight elements of ST were rated as "very helpful" across all versions of ChatGPT. The same was true for the corresponding questions in the case studies, with ChatGPT effectively relating each topic to the elements. With one minor exception being the version "GPT 4 + Show Me," which provided more generic answers and was consequently considered only as "helpful." This is illustrated by two different answers in respect to macroeconomic policy and the ST-element "Unterstanding Systems at Different Scales":

> Understanding Systems at Different Scales: Macroeconomic policy operates at different scales, from the national level to the global level. Policies designed at the national level can have effects beyond borders, such as through international trade, capital flows, and exchange rates. Understanding the interconnections and spillover effects between different scales of the macroeconomic system is crucial for effective policy formulation and coordination. (GPT 3.5, p. 192)

> Understanding Systems at Different Scales: Macroeconomics looks at the economy on a large scale, at the level of the whole economy, while microeconomics looks at individual markets. Both perspectives are important for a full understanding of economic phenomena. (GPT 4 + "Show Me", p. 231)

RQ3: How and to what degree can ChatGPT help learning about CLDs and support creating them?

The first four questions from Appendix A, which pertain to general inquiries about CLDs, received a consistently high rating of 'very helpful' in all versions. However, when participants were prompted to provide an example of a CLD (Question 5), ChatGPT's performance varied. GPT 3.5 received a rating of "partially helpful, " GPT 4 was rated as "helpful," and "GPT 4 + Show Me" was deemed "not helpful" as it failed to create a CLD altogether. The issue with the other two versions was related to the quality of the visualizations, which occasionally omitted the polarity or arrowheads. Nevertheless, the accompanying answers to the questions were accurate and the visualizations remained helpful. Notably, one of the two runs of GPT4 generated a reasonably good visualization (Figure 1).

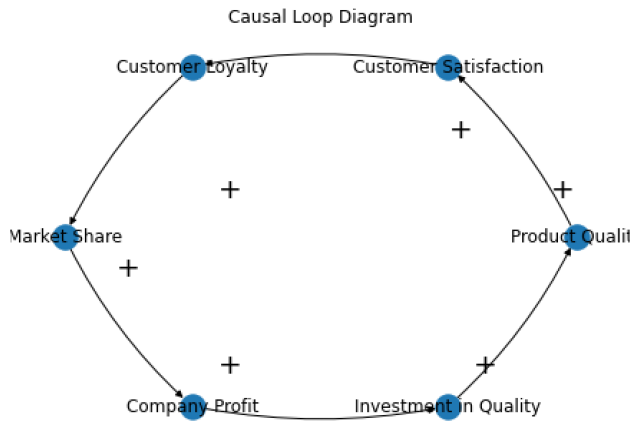

Figure 1: Causal Loop Diagram based on Python-code provided by GPT4 (p. 93)

The case studies presented a similar pattern: Overall, the identification and explanation of relevant elements and their connections were rated as very good and helpful, with "GPT 4 + Show Me" performing slightly worse. Notably, ChatGPT demonstrated a particularly useful ability to explain the connections. ChatGPT's ability to explain the connections was especially helpful. As previously mentioned, the quality of the visualizations varied randomly. However, when "Code Interpreter" was enabled for GPT 4, it produced high-quality visualizations (Figure 2).

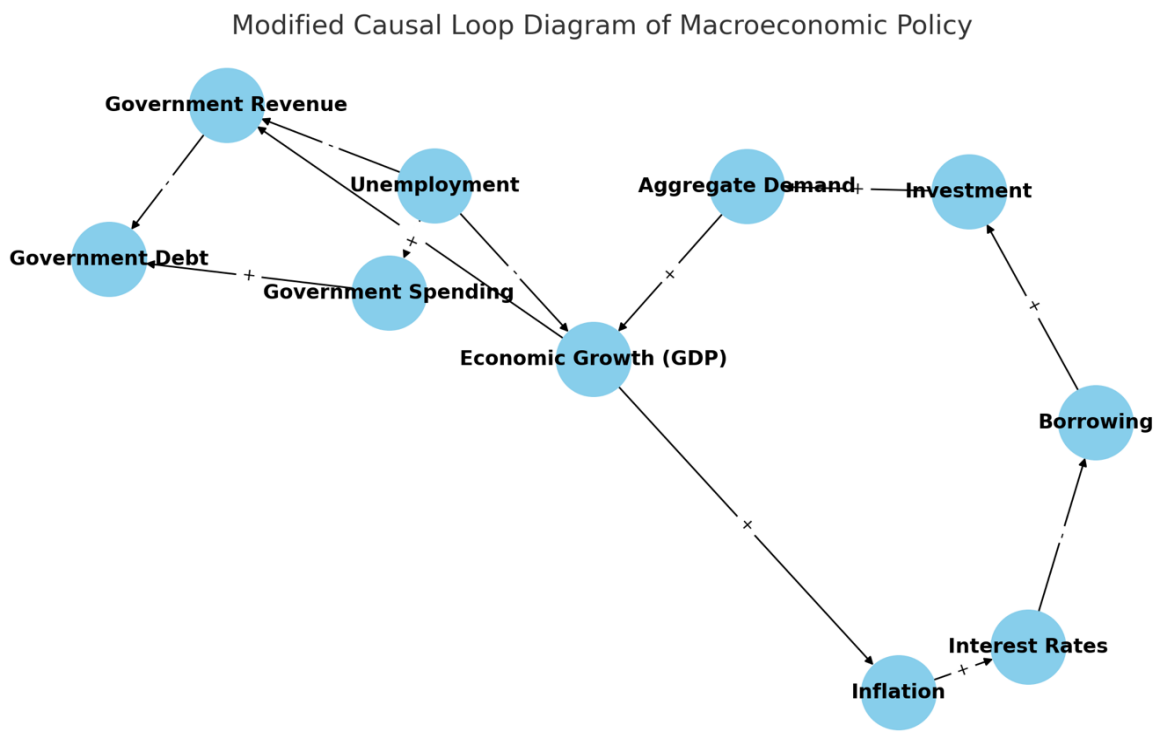

Figure 2: Causal Loop Diagram created by GPT4 with "Code Interpreter" (p. 250)

Often, ChatGPT doesn't generate an appealing visualization on the first attempt, but with persistence and specific guidance on improvements, the final result can be satisfactory (e.g., pp. 213-216).

In addition to the variability in visualization quality, the primary issue with CLDs, as described in RQ1, is the occasional incorrect polarity.

RQ4: How and to what degree can ChatGPT help learning about SD and support creating SD-models?

All five SD-questions from Appendix A received consistently high ratings of being correct and very helpful across all versions. This is evident in ChatGPT's exemplary answers during the case studies. The model demonstrated a remarkable ability to identify relevant stocks, flows, and variables, including their associated formulas. It excelled in answering detailed questions and readily accommodated user preferences in adjusting the model. Moreover, ChatGPT provided comprehensive, step-by-step instructions on implementing the model in various modeling software. However, some of the software-specific instructions were not entirely accurate, such as indicating the location of a particular option within a menu. Additionally, there was one instance where the step-by-step guidance proved to be excessively burdensome and, consequently, less useful (heat transfer with GPT 3.5, pp. 270-278).

RQ5: Are there any notable differences in quality between GPT3.5, GPT4, and GPT4 with plugins when utilized for acquiring ST-skills?

With the exception of "GPT4+Show Me," all versions consistently provided correct and highly helpful answers. Surprisingly, this was also true for GPT 3.5, which is a free version and was trained on a much smaller text corpus compared to GPT 4.

Both GPT 3.5 and GPT 4 demonstrated their capability to generate Python code, facilitating the creation of CLDs and simulation runs. The potential advantage of the "Show Me" plugin for CLDs could only be fully realized after multiple interactions to guide the tool accurately, if it proved helpful at all. Given the reduced quality of its answers, the usage of ChatGPT with plugins should be reserved for specific cases that genuinely benefit from its limited visualization capabilities.

The recently introduced "Code Interpreter" performed the given tasks in the case studies with the same quality as plain GPT 4. Its ability to directly run Python code eliminated the need for external runtime environments, providing a much more convenient experience. Additionally, it performed well in creating well-visualized CLDs.

RQ6: Are the answers provided by ChatGPT reliable?

When asking ChatGPT the same question twice to assess the reliability of its answers (within a narrow time frame of a couple of days to account for potential changes in ChatGPT's programming and resulting quality differences), only minimal differences were identified. In most cases, the explanations were very similar (26 times), or the variations involved slightly different content, primarily due to the use of different examples (38 times). This variation should not be seen as a problem; instead, it can be viewed as an advantage, as users can request different examples, gaining a better understanding of the topic. Only two instances showed a relevant difference in quality, specifically concerning the example of a CLD (p. 85 and p. 89.

**Discussion**

The results of the study demonstrate that ChatGPT performed remarkably well in the areas under investigation. The explanations, examples, and modeling guidance provided were largely accurate and helpful. This held true across various topics and subjects, spanning general aspects of systems thinking (ST) and System Dynamics (SD) modeling, including detailed step-by-step instructions, which are particularly valuable for novices. Despite occasional errors in creating CLDs and getting the polarities wrong, ChatGPT remains highly helpful in identifying relevant elements and connections, as well as explaining the underlying logic behind its recommendations.

The results also demonstrate that ChatGPT consistently maintains the quality of its answers when used within a similar time frame. This predictability is beneficial for classroom use, as students can expect similar results when employing appropriate prompts. However, it's crucial to recognize that variations in the way prompts are formulated can yield significantly different answers. Addressing this issue in class becomes essential, as it allows students to develop an understanding of the importance of well-constructed prompts and acquire corresponding expertise. Instructors can support this learning process by providing students with prompts to experiment with (as demonstrated in the appendices) or by comparing different prompts and the resulting answers.

When considering which GPT version to use, the free GPT3.5 performed remarkably well. Therefore, if budget constraints are a concern, GPT3.5 could be sufficient for most situations. On the other hand, if GPT4 is available, it is recommended to use it with 'Code Interpreter' due to its additional functionalities, such as creating visualizations or running simulations within its own environment. However, the use of other plugins, such as 'Show Me,' should be approached with caution, as it may lead to a significant deterioration in answer quality. It's important to note that these recommendations are based on data collected from May to July, and the quality of the versions may change over time.

Due to the possibility of unpredictable quality changes and occasional incorrect answers, both teachers and students should approach ChatGPT's responses critically. If something seems unclear or doesn't make sense, it is essential to seek clarification or more details from ChatGPT. If the answers remain unconvincing, the students should not hesitate to consult their teacher, or other alternative sources. This critical approach to information evaluation is not specific to learning with ChatGPT but is an integral part of fostering critical thinking, which teachers should aim to cultivate.

With sufficient emphasis on critical thinking, ChatGPT can serve as a valuable tool for enhancing ST- skills, enabling highly interactive and personalized learning experiences. Students can request adjusted difficulty levels, access numerous examples, and seek varying levels of detail in explanations. Additionally, they can ask questions they might be hesitant to ask a teacher or demand further explanations as needed. As an interactive tool providing instant responses, ChatGPT has the potential to stimulate student curiosity and foster self-directed learning.

Based on the results of this study, ChatGPT can indeed function as a personal tutor and contribute to the enhancement of ST skills.

In terms of study **limitations**, it is important to acknowledge that the quality of ChatGPT's answers appears to vary over time. Thus, its good performance in this study may not be consistently replicated, and its utility as a learning aid might differ from what the results imply.

Furthermore, the analysis focused on only a limited selection of ST aspects and subjects. Different topics and subjects may yield less favorable results. Additionally, the study employed a novice perspective when working with ChatGPT. An expert might pose more in-depth questions, potentially reaching the limits of ChatGPT more frequently.

The study predominantly utilized prompts listed in appendices A and B. While slight variations in prompts can be expected in real classroom situations, they may significantly impact the answer quality.

Lastly, it is important to note that the study has an explorative character, with a primary focus on theoretical analysis. It does not empirically test hypotheses in real-world learning situations. Conducting such empirical evaluations, especially regarding the helpfulness of ChatGPT's answers, would be highly valuable and interesting.

Given the limitations of this study, **future research** is necessary to monitor potential changes in the answer quality over time. It would be interesting to explore how ChatGPT performs in different topics to identify potential weak areas. As this study primarily adopted a novice perspective, it is essential for other studies to evaluate ChatGPT's usefulness for experts, including its capabilities in modeling complex situations in depth.

To broaden the perspective, it would be valuable to evaluate other AI tools as well. In this context, research should also focus on the customization of AI tools to better cater to specific educational contexts. For instance, future iterations of AI models could be trained on specific educational datasets to provide more accurate and tailored information in a given subject, such as systems thinking.

Empirical studies are needed to evaluate the actual impact of integrating AI tools into education on students' learning outcomes. Rigorous experimental designs, such as randomized control trials, could provide valuable insights into the effectiveness of this teaching method and allow researchers to isolate and study the influence of various factors.

As the role of the teacher evolves from being the primary source of knowledge to a facilitator and guide in the AI-aided learning process, further research is required to understand how this transformation impacts teaching practices and the overall dynamics in the classroom. Qualitative studies exploring teachers' experiences, challenges, and strategies when integrating ChatGPT into their lessons would yield valuable information.

Long-term studies are necessary to assess the sustained impact of AI integration on students' systems thinking skills and other educational outcomes, including emotional and motivational aspects. Such studies could provide insights into how continued interaction with AI tools like ChatGPT shapes students' learning trajectories, problem-solving abilities, and readiness for the complex, interconnected world.

**Appendix A – Prompts used for general learning about systems thinking**

**Basic Concepts / General orientation**

1. What is systems thinking?
2. Why is systems thinking important in today's world?
3. What are the main elements of a system?
4. Can you provide an {**subject**, e.g., **economic**/biological/physical} example of a system?

**Elements of ST**

1. Recognizing interconnections is considered to be an important element of systems thinking. Explain this concept. Give me an example.

2. Identifying and understanding feedback is considered to be an important element of systems thinking. Explain this concept. Give me an example.

3. Understanding system structure is considered to be an important element of systems thinking. Explain this concept. Give me an example.

4. Differentiating types of stocks, flows and variables is considered to be an important element of systems thinking. Explain this concept. Give me an example.

5. Identifying and understanding non-linear relationships is considered to be an important element of systems thinking. Explain this concept. Give me an example.

6. Understanding dynamic behavior s considered to be an important element of systems thinking. Explain this concept. Give me an example.

7. Reducing complexity by modeling systems conceptually is considered to be an important element of systems thinking. Explain this concept. Give me an example.

8. Understanding systems at different scales is considered to be an important element of systems thinking. Explain this concept. Give me an example.

**Causal Loop Diagrams (CLD)**

1. What is a causal loop diagram?

2. How can a causal loop diagram help visualize a system?

3. Can you explain the process of creating a causal loop diagram?

4. What challenges might one face when creating a causal loop diagram?

5. Create an example of a causal loop diagram (CLD) {using Python}. It should have relevant elements connected with arrows and a + or - to indicate the correlation. Also, name the relevant elements, list all connections including + and –, and explain them.

**System Dynamics (SD)**

1. What do I need to know when I want to build a system dynamics model? – with follow-up questions if needed

2. What are stocks, flows and variables in systems dynamics? Provide an example.

3. What is meant by "delay" in systems dynamics? Provide an example.

4. How can a delay influence system behavior?

5. I want to learn how to build a system dynamics model. Show me how to proceed with a simple example (you chose the topic): ask relevant questions, answer them yourself and build the model. Also provide code so I can run it in Python.

**Appendix B – Standard questions used in the in-depths analysis**

*Initial prompt for general information about the topic with a subject-specific focus:* You're an {economist} and an expert for systems thinking. I would like to learn something about {macroeconomic policy (especially in respect to …)} and take the perspective of an {economist}. Start by giving me an overview of the topic.

*Relate the topic to the 8 elements of systems thinking:* I would also like to improve my systems thinking skills. Accordingly, I want to relate the topic of Macroeconomic policy to these elements of systemic thinking: recognizing interconnections, identifying and understanding feedback, understanding system structure, differentiating types of stocks, flows and variables, identifying and understanding non-linear relationships, understanding dynamic behavior, reducing complexity by modeling systems conceptually and understanding systems at different scales.

*CLD Structure:* Now I'd like to create a causal loop diagram (CLD) to visualize important connections. It should have relevant elements connected with arrows and a + or - to indicate the correlation. 1. Tell me which elements to use 2. Tell me how to connect them with arrows. Always tell if + or -. And always explain why.

*CLD-feedback loops:* List and explain all feedback loops in that CLD-system.

*CLD visualization:* Create this CLD. If you can't do it otherwise, create Python code.

*SD-modelling content:* Now let's proceed to creating a system dynamics model. {*Depending on the subject's complexity:* Let's start with only a part of what we modeled in the CLD.} What would you propose to begin with?

*SD-model structure:* Which stock, flows and variables would I need? We also need formulas and values to create the model.

*Step-by-step-modelling:* Finally, I'd like to create this model in a software tool like Insight Maker. Tell me step by step how to proceed.

*Do a simulation run (creates Python Code, except when using "Code Interpreter):* Run the simulation.